# EmoTxt: A Toolkit for Emotion Recognition from Text

Fabio Calefato, Filippo Lanubile, Nicole Novielli
*University of Bari "Aldo Moro"*
{fabio.calefato,filippo.lanubile,nicole.novielli}@uniba.it

*Abstract*— We present EmoTxt, a toolkit for emotion recognition from text, trained and tested on a gold standard of about 9K question, answers, and comments from online interactions. We provide empirical evidence of the performance of EmoTxt. To the best of our knowledge, EmoTxt is the first open-source toolkit supporting both emotion recognition from text and training of custom emotion classification models.

## 1. Introduction

Sentiment analysis is regarded as a crucial task for several application domains, including business, social well-being, politics, security, and software engineering [1]. To date, several off-the-shelf tools are freely available for classifying the sentiment polarity of an input text, that is its *positive*, *negative*, or *neutral* semantic orientation [14]. However, none of them supports the recognition of specific emotions, such as joy, love, and anger. In this paper, we present EmoTxt, the first open-source toolkit for emotion recognition from text. The toolkit can be used by researchers for detecting emotions from input text as well as for training a custom emotion classifier from scratch, based on manually annotated data.

## 2. System Description

The system is completely developed in Java and distributed under the MIT open-source license[1]. With the toolkit, we release the classification models trained on our gold standard datasets (see Section 3.1), which can be used for emotion detection from text. EmoTxt identifies emotions in an input corpus provided as a comma separated value (CSV) file, with one text per line, preceded by a unique identifier. The output is a CSV file containing the text id and the predicted label for each item of the input collection.

EmoTxt can also be used to train a new classifier starting from a gold standard dataset with custom emotion labels. To train a new emotion classifier, a collection of texts with hand-provided emotion labels should be provided. The training approach implemented by EmoTxt is independent of the classification framework and learns how to detect the presence/absence of a given emotion based on the labels manually provided in the training examples. Detailed guidelines on how to use EmoTxt are available at https://goo.gl/Mjd6y2.

The system is based on the approach described by Ortu et al. [9]. In addition to **uni-** and **bi-grams**, modeled using a *tf-idf* weighting schema, we leverage a suite of lexical features capturing the presence of lexical cues conveying emotions in the input text:

i. **Emotion Lexicon**: we capture the presence of emotion lexicon by relying on the association between words and emotion categories in WordNet Affect [12]. In particular, we compute the *tf-idf* for each emotion category (e.g., joy, love, enthusiasm, sadness, ingratitude, etc.) based on the occurrences of words associated with them category it in WordNet Affect;
ii. **Politeness** conveyed by the text and measured using the tool developed by Danescu et al. [2];
iii. **Positive** and **Negative Sentiment Scores**, computed using SentiStrength [14], a publicly available tool for sentiment analysis;
iv. **Uncertainty**, as measured by the tool of De Smedt et al. [3] based on the of grammatical moods and adverbs uncertainty.

Before proceeding to feature extraction, we tokenize texts using the Stanford NLP library[2]. Also, we removed HTML tags, code fragments, and URL that might introduce noise in the training. Unlike Ortu et al. [9], we did not perform any stemming nor lemmatization as inflected forms may convey important information about sentiment.

## 3. Evaluation

### 3.1. Dataset

Mining emotions from text requires choosing the most appropriate model to operationalize sentiment. In this study, we used two gold standards datasets annotated according to the discrete framework by Shaver et al. [11]. The framework defines a tree-structured hierarchical classification of emotions, where each level refines the granularity of the previous one, thus providing more indication on its nature. The framework includes, at the top level, six basic emotions, namely *love*, *joy*, *anger*, *sadness*, *fear*, and *surprise*. The framework is easy to understand, thanks to the intuitive nature of the emotion labels. It has already been used for emotion mining in software engineering [5][9].

To train and evaluate our classifier for emotions, we built a gold standard composed of 4,800 posts (question, answer, and comments) from Stack Overflow (SO), an online community where over 7 million programmers do networking by reading and answering others' questions, thus participating in the creation and diffusion of crowdsourced knowledge and software documentation[3]. The SO gold standard was annotated by 12 raters. The annotation sample was extracted from the official SO dump of user-contributed content from July 2008 to September 2015. The coders were requested to indicate the presence/absence of each of the six basic emotions from the Shaver framework. Each post was annotated by three raters and we resolved the disagreements by applying a majority voting criterion. The observed agreement, ranging from .86 for joy to .98 for surprise, confirms the reliability of our gold standard.

As a further evaluation, we also assessed the performance of

---

[1] EmoTxt is available for download at: https://github.com/collab-uniba/Emotion_and_Polarity_SO

[2] https://stanfordnlp.github.io/CoreNLP/download.html
[3] https://stackoverflow.com/ Last accessed: July 2017.

EmoTxt on the gold standard released by Ortu et al. [9]. Their dataset includes 4000 comments posted by software developers on Jira[4], one of the most popular issue tracking systems among software companies. The Jira gold standard contains sentences manually labeled with the emotions *love*, *joy*, *anger,* and *sadness* (see TABLE I).

TABLE I.    EMOTION LABEL DISTRIBUTION IN THE TWO DATASETS.

| Dataset | Texts conveying the emotion | | | | | | N |
|---|---|---|---|---|---|---|---|
| | *Love* | *Joy* | *Surprise* | *Anger* | *Sadness* | *Fear* | |
| SO | 1220 | 491 | 45 | 882 | 230 | 106 | 4800 |
| Jira | 166 | 124 | *NA* | 324 | 302 | *NA* | 4000 |

We split both gold sets into training (70%) and test (30%) partitions using the R package caret [8] for stratified sampling. We use the training set to seek the optimal parameter setting for our classifier. For each dataset, the final model is trained on the training set using the optimal configuration and then evaluated on the test set, to assess the model performance on unseen new data from the held-out test set (see Section 3.2).

### 3.2. Model Training and Classification Performance

We trained the EmoTxt classification models in a supervised machine learning setting using Support Vector Machines (SVM). In particular, we used the R interface [6] to Liblinear [4], an open source library for large-scale linear classification with SVM. Linear SVM is a state-of-the-art learning technique for such high-dimensional sparse datasets with a large number of items and a large number of features $N$, where each item has only $s \ll N$ non-null features [7], as typical in presence of n-grams. A classifier performance strongly depends on the setting of its input parameters [13]. For linear classification, the only parameter is the cost parameter C. Too large C values makes the cost of misclassification high, thus forcing the algorithm to better explain the training data but potentially inducing the risk of overfitting. To fine-tune the SVM parameter while still preventing overfitting, we ran the Liblinear parameter tuning utility on our training set in a 10-fold cross-validation setting over the train partition of our datasets. We chose the optimal value for the C parameter by maximizing the prediction accuracy, searching for C values in {0.01, 0.05, 0.10, 0.20, 0.25, 0.50, 1, 2, 4, 8}. We trained EmoTxt as a suite of six binary classifiers, which predict the presence/absence of each emotions in the input text. The performance on the test set is reported in TABLE II.

TABLE II.    CLASSIFIER PERFORMANCE.

| | Stack Overflow | | | Jira | | |
|---|---|---|---|---|---|---|
| Emotion | Prec | Rec | F1 | Prec | Rec | F1 |
| Joy | 0.77 | 0.77 | 0.77 | 0.85 | 0.85 | 0.85 |
| Love | 0.92 | 0.92 | 0.92 | 0.86 | 0.86 | 0.86 |
| Sadness | 0.79 | 0.79 | 0.79 | 0.83 | 0.83 | 0.83 |
| Anger | 0.86 | 0.86 | 0.86 | 0.75 | 0.74 | 0.74 |
| Surprise | 0.58 | 0.58 | 0.58 | | | |
| Fear | 0.86 | 0.86 | 0.86 | | | |

### 4. Conclusions

We present EmoTxt, an open source toolkit for emotion detection from text, trained and tested on two large gold standard datasets mined from Stack Overflow and Jira. We release the classification models to be used for emotion detection tasks. Other than classification, EmoTxt supports training of emotion classifiers from manually annotated training data. Its training approach leverages a suite of features that are independent of the theoretical model adopted for labeling the data. We provide empirical evidence that EmoTxt achieves comparable performance with different datasets. As future work, we plan to validate it on gold standards from different sources, to further assess the generality and robustness of the approach implemented.

### Acknowledgments

This work is partially supported by the project 'EmoQuest', funded by the Italian Ministry of Education, University and Research under the SIR program.

---

[4] https://www.atlassian.com/software/jira